\documentclass[useAMS, usenatbib]{mn2e}

\usepackage{rotating}
\usepackage{natbib}
\usepackage{latexsym}
\usepackage{amssymb}
\bibpunct{(}{)}{;}{a}{}{,} 

\def\hide#1{}

\def \apj {ApJ}

\def \apjl {ApJL}
\def \aap {A\&A}

\begin{document}

\author[Linares et al.]{\parbox[t]{\textwidth}{\raggedright Manuel
  Linares$^1$\thanks{linares@uva.nl}, Anna Watts$^1$, Rudy
  Wijnands$^1$, Paolo Soleri$^1$, Nathalie Degenaar$^1$, Peter A.
  Curran$^1$, Rhaana L. C. Starling$^2$ and Michiel van der
  Klis$^1$}\\
\vspace*{6pt}\\
$^1$ Astronomical Institute ``Anton Pannekoek'', University of
Amsterdam and Center for High-Energy Astrophysics, 
\\ Kruislaan 403,
NL-1098 SJ Amsterdam, Netherlands 
\\ $^2$ Department of Physics and Astronomy, University of Leicester, 
 University Road, Leicester LE1 7RH, UK }

\title[Long burst XTE~J1701--407]{The {\it Swift} capture of a long X-ray burst from XTE~J1701--407}


\maketitle

\begin{abstract}

XTE~J1701--407 is a new transient X-ray source discovered on June 8th,
2008. More than one month later it showed a rare type of thermonuclear
explosion: a long type I X-ray burst. We report herein the results of
our study of the spectral and flux evolution during this burst, as
well as the analysis of the outburst in which it took place. We find
an upper limit on the distance to the source of 6.1 kpc by considering
the maximum luminosity reached by the burst. We measure a total
fluence of 3.5$\times 10^{-6}$ erg/cm$^2$ throughout the $\sim$20
minutes burst duration and a fluence of 2.6$\times 10^{-3}$ erg/cm$^2$
during the first two months of the outburst. We show that the flux
decay is best fitted by a power law (index $\sim$1.6) along the tail
of the burst. Finally, we discuss the implications of the long burst
properties, and the presence of a second and shorter burst detected by
{\it Swift} ten days later, for the composition of the accreted material and
the heating of the burning layer.

\end{abstract}


\section{Introduction}
\label{sec:intro}

The bulge of our Galaxy harbours a large fraction of the known
population of accreting compact objects. During one of the regular
monitoring observations of that region with the proportional counter
array (PCA) onboard the Rossi X-ray timing explorer ({\it RXTE}) on June
8th, 2008, a new transient X-ray source was discovered: XTE~J1701--407
\citep{Markwardt08}. Three days later a follow-up observation with the
X-ray telescope (XRT) onboard {\it Swift} improved the source localization
and showed a spectrum that could be fitted with an absorbed power law
of index $\sim$2 \citep{Degenaar08}. The transient nature, flux level
and spectrum of the source are characteristic of X-ray binaries, but
the exact class of the binary (high-mass or low-mass) was unkown and
the compact object (black hole or neutron star) remained unidentified
for more than one month. On July 17th at 13:29:59 UT the burst alert
telescope (BAT) onboard {\it Swift} detected an X-ray flare at a position
consistent with that of XTE~J1701--407 \citep{Barthelmy08}. {\it Swift}
began slewing to the source 60 seconds after the trigger and the XRT
began observing the field 97 seconds after the initial BAT trigger,
detecting a fading X-ray source at the position of XTE~J1701--407
\citep{Barthelmy08}. \citet{Markwardt08b} suggested that the flare
could be caused by a thermonuclear burst based on the early BAT
data. By studying the evolution of the XRT spectrum, \citet{Linares08}
found a clear cooling curve and confirmed the speculation that this
was a thermonuclear event, identifying the source as an accreting
neutron star (NS) and the system, in all likelihood, as a low-mass
X-ray binary (LMXB). As noted by \citet{Linares08} this particular
event, having a total duration of more than 15 minutes, belongs to the
rare subclass of long duration bursts \citep{Cumming06}.

After the long X-ray burst the source continued in
outburst\footnote{To clarify the terminology we stress that in the
context of LMXBs an outburst is powered by accretion and lasts for
weeks to years whereas a type I X-ray burst, or ``burst'', results
from unstable thermonuclear burning on the surface of a NS and lasts
for $\sim$10 seconds to $\sim$hours.} and {\it RXTE} kept observing it. A
refined {\it Swift}-XRT position \citep{Starling08}, a search for an IR
counterpart \citep{Kaplan08} and the discovery of kHz quasi-periodic
oscillations \citep[QPOs;][]{Strohmayer08} were all reported during
the following two weeks. On July 27th BAT detected another X-ray flare
from XTE~J1701--407 \citep{Sakamoto08}. The soft spectrum and duration
($\sim$10~s) of the flare indicated that this was a type I X-ray
burst (http://gcn.gsfc.nasa.gov/notices\_s/318166/BA/), and
hereafter we refer to it as the short burst (even though from the BAT
data the total duration is somewhat uncertain). At the moment of
writing the source is still active, more than two months after the
start of the outburst.

In this Letter we present our analysis of the bursting behaviour of
this source. Long bursts are rare, and therefore probe quite unusual
burning regimes. Their duration is between that of normal Type I X-ray
bursts (durations $\sim 10-100$ s, triggered by unstable burning of
H/He) and superbursts (durations $\sim$ hours, triggered by unstable
burning of C). There are two scenarios in which such bursts could
occur, both requiring the build-up and ignition of a thick layer of
He. One possibility is that the system is ultracompact, so that it
accretes nearly pure He \citep{Cumming06}. The other possibility is
that unstable H burning at low accretion rates builds up He
\citep{Peng07, Cooper07}. While several examples of the first type are
now known (see for example \citealt{intZand07, Falanga08} and
references therein), the second class - which probe the boundaries of
H burning stability - have so far proved much rarer
\citep{Chenevez07}. In this Letter we will demonstrate that XTE
J1701-407 is a good candidate for membership of this second class.

\section{Observations and Data Analysis}
\label{sec:data}

We analysed all the 27 pointed {\it RXTE}-PCA observations of
XTE~J1701--407 taken until August 10th, 2008 (ID 93444-01). We also
analysed the {\it Swift}-XRT data taken during and after the long
burst (on July 17th, obsids 00317205000 and 00317205001). {\it Swift}
observed XTE~J1701--407 again on July 27th, triggered by the BAT
detection of the short burst (Sec.~\ref{sec:intro}). We also included
this XRT observation (obsid 00318166000) in our analysis, which
started $\sim$109 seconds after the trigger but did not detect the
burst. In the following two sections we provide the details of our
analysis. All errors given in Section~\ref{sec:results} correspond to
the 1~$\sigma$ confidence level.

\subsection{Swift}
\label{sec:dataswift}

We ran the XRT pipeline (v. 0.12.0) on all the {\it Swift}
observations. For the spectral analysis we created exposure maps and
ancilliary files and used the latest response matrices (v011) provided
by the {\it Swift}-XRT team. We grouped the spectra to a minimum of 20
counts per bin, applyied a systematic error of 2.5\% \citep{Campana08}
and fitted the 0.5-10~keV spectra within Xspec
\citep[][v. 11]{Arnaud96}. In our fits (see models below) we fixed the
absorbing column density to the average value found from X-ray fits to
the persistent emission: 3.5$\times 10^{22}$ cm$^{-2}$
\citep[][]{Degenaar08,Markwardt08}.

Observation 00317205000 started on July 17th, 2008 at 13:31:36.9 UT
(Sec.~\ref{sec:intro}) and lasted for a total of about one
kilosecond. All the XRT data were collected in windowed timing (WT)
mode, except for the initial $\sim$8 seconds which were taken in
photon counting (PC) mode. We divided the observation into nine
contiguous intervals of increasing length in order to compensate for
the flux decay and maintain an approximately constant total number of
counts in each interval (between $\sim$1500 and $\sim$2500). About 830
seconds after the start of the observation a data gap of $\sim$3300
seconds occurred, after which the XRT continued to observe the source
for another $\sim$35 seconds. This last data segment allows us to
estimate the persistent flux after the X-ray burst. We extracted
source and background spectra for each interval using circular regions
with radii of $\sim$30 and $\sim$15 pixels, respectively, and fitted
the resulting spectra using an absorbed black body model (reduced
$\chi^2$ between 0.7 and 1.1).

We performed fast Fourier transforms on 2s segments of XRT-WT data
taken during the long burst in order to search for rapid
variability. For that purpose we used the energy range 0.5-10 keV,
keeping the original time resolution and thereby sampling the
$\sim$0.5-280~Hz frequency range.

Both post-burst observations (00317205001 and 00318166000) allow us to
characterize the persistent soft X-ray emission of XTE~J1701--407. For
the PC data we used an annular extraction region with radii 10-25
pixels in order to account for pile-up. The energy fraction enclosed
by the resulting annuli was about 20\%, and such energy loss was taken
into account when creating the ancilliary files. Finally, including
also the WT data, we fitted the energy spectra with an absorbed power
law model (reduced $\chi^2$ of 1.1 and 1.2).

\subsection{RXTE}
\label{sec:datarxte}

We analysed all the pointed {\it RXTE} observations of the source made
until August 10th. After applying standard filters we extracted energy
spectra from Standard 2 data collected by PCU2 and corrected them for
background using the latest PCA background models. We generated
reponse matrices and fitted the spectra within Xspec in the 3.0-25.0
keV band, after applying a 1\% systematic error, with a black body
plus power law model corrected for absorption (see
Sec.~\ref{sec:dataswift}). The resulting fits had reduced $\chi^2$
between 0.5 and 1.4 (for 45 degrees of freedom). We extracted a
hardness ratio (``hard colour'' hereafter) for each observation,
defined as the count rate in the 9.7-16.0 keV band divided by that in
the 6.0-9.7 keV band and normalized by that of the Crab. Furthermore,
we extracted a background and deadtime corrected HEXTE (cluster B)
energy spectrum from each observation, but found no significant source
emission above 50 keV in any of them.

\section{Results}
\label{sec:results}

\subsection{Outburst}
\label{sec:outburst}

\begin{figure}
  \begin{center}
  \resizebox{1.0\columnwidth}{!}{\rotatebox{-90}{\includegraphics[]{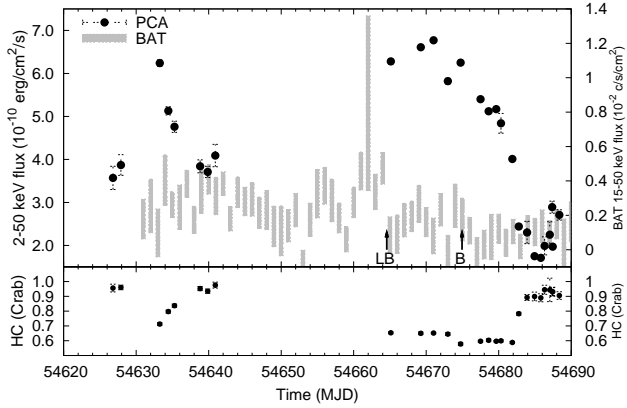}}}
  \caption{Time evolution of the 2-50 keV unabsorbed flux {\it (top, filled dots)} and the hard colour {\it (bottom)} during the outburst of XTE~J1701--407, as measured by the {\it RXTE}-PCA. Arrows indicate the times of the long (LB) and short (B) X-ray burst. The 15-50 keV flux measured by {\it Swift}-BAT is shown as well {\it (top, grey rectangles)}.}
    \label{fig:lc}
 \end{center}
\end{figure}

The overall lightcurve is shown in Figure~\ref{fig:lc}, together with
the evolution of the hard colour. The outburst shows two peaks at a
similar flux, (6.2, 6.8)$\times 10^{-10}$ erg/cm$^2$/s. The BAT
transient monitor
lightcurves\footnote{http://swift.gsfc.nasa.gov/docs/swift/results/transients/}
indicate that the source was active during the three weeks where no
{\it RXTE} pointed observations were obtained, at a flux level similar
to that measured during the first two weeks of outburst (see
Fig.~\ref{fig:lc}). The hard colour traces directly the changes in
spectral state along the outburst, with a clear tendency: when the
source is bright its spectrum is soft whereas the spectrum is harder
at the lowest fluxes. However, it is interesting to note that as is
the case for many other NS-LMXBs the correspondence between flux and
hardness is not one-to-one \citep{vanderKlis06}. Together with the
presence of kHz quasi-periodic oscillations \citep{Strohmayer08}, the
peak luminosity and the spectral evolution (Fig.~\ref{fig:lc})
show that XTE~J1701--407 is a new member of the atoll
source class \citep{HK89}.  Both X-ray bursts occur at relatively high
fluxes (see Table~\ref{table:bursts}) and soft spectra (low hard
colour).

By interpolating linearly between the available datapoints we
integrate the lightcurve and find a total outburst fluence up until
August 10th of 2.6$\times 10^{-3}$ erg/cm$^2$. More than half of this
energy (1.6$\times 10^{-3}$ erg/cm$^2$) was radiated before the long
burst occurred, whereas the fluence between long and short bursts was
about 5$\times 10^{-4}$ erg/cm$^2$.

From our spectral fits to the post-burst XRT observations
(Sec.~\ref{sec:dataswift}) we measure 2-10~keV fluxes of
($4.0\pm0.9)\times 10^{-10}$ erg/cm$^2$/s (July 17th) and
($5.6\pm1.0)\times 10^{-10}$ erg/cm$^2$/s (July 27th), which are
consistent with the fluxes in the same band obtained from the nearest
(within one day) PCA observations. In both cases the photon index was
$\sim$2. Furthermore, we estimate unabsorbed 1-2~keV fluxes of
$2.3\times 10^{-10}$ erg/cm$^2$/s (July 17th) and $3.5\times 10^{-10}$
erg/cm$^2$/s (July 27th) and add them to the unabsorbed 2-50~keV
fluxes measured with the PCA within one day of the bursts ($6.3\times
10^{-10}$ erg/cm$^2$/s for both bursts; see Fig.~\ref{fig:lc}). These
1-50~keV fluxes (Table~\ref{table:bursts}) are the closest we can
safely get to the bolometric persistent flux at the times of the
bursts, but we note that the uncertainties in the absorbing column
density and the extrapolation to the 0.01-1~keV energy range could
change the bolometric flux by a factor of $\sim$2 \citep[see
e.g. discussion in][]{intZand07}.

\subsection{X-ray bursts}
\label{sec:burst}

From the flux and spectral parameters reported by \citet{Markwardt08b}
we estimate a bolometric unabsorbed peak flux of 8.4$\times 10^{-8}$
erg/cm$^2$/s for the long burst. By assuming that the peak luminosity
was equal to or lower than that typical of photospheric expansion type
I X-ray bursts \citep[$\sim$3.8$\times 10^{38}$ erg/s][]{Kuulkers03}
we can place an upper limit on the distance to the source of 6.1
kpc. On the other hand, if we use the Eddington limit for a mix of H
and He ($\sim$1.6$\times 10^{38}$ erg/s for a hydrogen fraction X=0.7)
as a limit for the peak luminosity the upper limit on the distance
becomes 4.0 kpc. The rise time in the BAT lightcurve was about 50
seconds, the peak occurred about ten seconds after the
trigger (http://gcn.gsfc.nasa.gov/notices\_s/317205/BA/) and
the total duration, assuming that the burst finished when it reached
1\% of the peak flux, was about 21 min. The fluxes in the post-burst
XRT observation, taken more than two hours after the peak of the long
burst, are similar to the flux we measure in the last interval
($\sim$1 hour after the peak, Sec.~\ref{sec:dataswift}), confirming
that the source had returned to the persistent level.

\begin{figure}
  \begin{center}
  \resizebox{0.6\columnwidth}{!}{\rotatebox{0}{\includegraphics[]{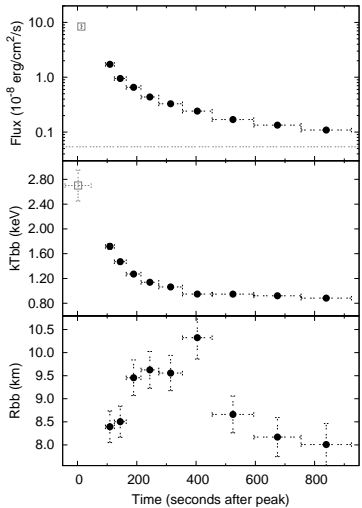}}} 
  \caption{Time evolution of the bolometric flux {\it(top)}, black body temperature {\it(middle)} and black body radius {\it(bottom)} along the long X-ray burst. Filled dots correspond to our {\it Swift}-XRT measurements whereas empty squares show {\it Swift}-BAT results \citep{Markwardt08b}. The horizontal dotted line on the top panel indicates the persistent flux. Error bars in the x axis represent the averaged intervals. The distance upper limit of 6.1 kpc has been used to obtain the black body radius.}
    \label{fig:burst}
 \end{center}
\end{figure}

Figure~\ref{fig:burst} shows the results of our time-resolved
spectroscopy of the tail of the long burst. The flux decays by almost
two orders of magnitude and the temperature decreases from
1.72$\pm$0.05~keV to 0.88$\pm$0.02~keV during the burst tail \citep[see
also][]{Linares08}. XRT did not observe the peak of the burst, and no
spectral variations were seen in BAT data of that period
\citep{Markwardt08b}. Therefore the question of whether or not the
burst showed photospheric radius expansion cannot be addressed. Going
back to our XRT measurements, the radius of the black body appears to
increase between $\sim$200~s and $\sim$500~s after the peak
(Fig.~\ref{fig:burst}). However, the large error affecting this
parameter, the uncertainties on the underlying model (NS atmosphere in
a dramatically-non-steady state) and the absence of an associated drop
in the black body temperature imply that this cannot be taken as
concrete evidence of a physical increase in emitting area.

\begin{figure}
  \begin{center}
  \resizebox{0.6\columnwidth}{!}{\rotatebox{0}{\includegraphics[]{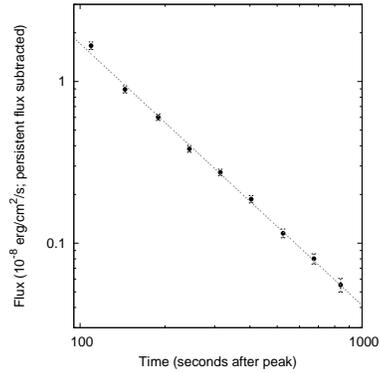}}}
  \caption{Fit to the long burst flux decay with a power law of index $\sim$1.6, together with the values of the baseline-subtracted bolometric flux.}
    \label{fig:decay}
 \end{center}
\end{figure}

We fit the flux decay of the long burst measured by the XRT and find
that a simple power law with an index of -1.62$\pm$0.04 and
normalization (3.0$\pm$0.6)$\times 10^{-5}$ erg/cm$^2$/s gives a
remarkably good description of the data (reduced $\chi^2$ of 1.2;
see Fig.~\ref{fig:decay}). A simple or double exponential decay do not
fit the data satisfactorily (reduced $\chi^2$ of 22 and 7,
respectively). We note that power law cooling is consistent with the
expectations from models for long bursts \citep{Cumming04}.

The bolometric fluence measured by XRT during the long burst, after
subtracting the persistent flux level, is 2.2$\times 10^{-6}$
erg/cm$^2$. We integrate the flux decay along the data gap
(Sec.~\ref{sec:dataswift}) and estimate a fluence of 4.4$\times
10^{-7}$ erg/cm$^2$ during that period. Adding the reported BAT
fluence \citep{Markwardt08b} we find a total burst fluence of
3.5$\times 10^{-6}$ erg/cm$^2$, which using the 6.1 kpc upper limit on
the distance corresponds to a maximum radiated energy of
1.6$\times 10^{40}$ erg.

We searched the XRT power spectra obtained during the tail of the long
burst (Sec.~\ref{sec:dataswift}) and found no strong pulsations or
QPOs. This non-detection is not surprising given i) the distribution
of burst oscillation frequencies \citep{Galloway06} combined with our
Nyquist frequency ($\sim$280 Hz) and ii) the relatively low count rate
collected by the XRT: for the maximum rate of $\sim$90 c/s, a
3$\sigma$ detection of a 2~Hz wide QPO in a 2s FFT would require a
fractional rms of at least $\sim$ 25\% \citep[][]{vanderKlis95b}, much
higher than the values usually measured \citep{Galloway06}.

\section{Discussion}
\label{sec:discussion}

Theoretical modelling of long bursts suggests that
they involve the build-up and ignition of a thick layer of He. One
possibility is that the binary is ultracompact, so that the accreted
fuel is almost pure He.  At low accretion rates, and in the absence of
heating from steady H burning, the temperature remains low and ignition
can be delayed until a thick layer of He has built up
\citep{intZand05, Cumming06}. Although He burning is quick, the
cooling time is long because of the thickness of the layer in which
the heat is deposited. Long bursts can also be triggered in systems
accreting a mix of H and He if the source accretes at low rates, close
to the point where H burning stabilises \citep{Fujimoto81, Peng07,
Cooper07}.  In this case weak H flashes can build up He, which ignites
sporadically, resulting in a long burst.  Layer thickness again
results in a long cooling time: but the presence of H in the burning
mix may also prolong nuclear energy generation.

Using the measured fluxes $F$ from Table \ref{table:bursts} and the upper limit on the distance $d$ that we
obtained from the long burst, we infer an upper limit on the global accretion rate $\dot{M} \approx
4\pi d^2 R F/GM$ of
3.7$\times 10^{-10}$ $M_\odot$ yr$^{-1}$ (for a neutron star mass $M =
1.4 M_\odot$ and radius $R = 10$ km) around the times of the two bursts.
Assuming isotropy, this
corresponds to an upper limit on the local accretion rate $\dot{m}$ of 1.9$\times 10^{3}$ g/cm$^2$/s,
(2.5\% of the Eddington rate for an Eddington luminosity of 1.6$\times 10^{38}$ erg/s).
This accretion rate is
close to the rate where H burning is expected to transition
 to stability \citep{Fujimoto81, Peng07, Cooper07}, so we need to examine both scenarios for the
generation of the long burst.

The BAT rise time of the long burst, $\approx 50$ s, is much longer
than most of the pure He bursts examined by \citet{intZand07} and
\citet{Falanga08}. This points to it being a mixed H/He type burst,
triggered by H ignition, although we note that burst lightcurves can
change considerably depending on the energy range where they are observed
\citep[see][]{Chelovekov06,Molkov05}.

We next consider the ignition conditions. The long burst had a total
energy output of at most $E_b = 1.6\times 10^{40}$ ergs (6.1 kpc
distance).  The corresponding ignition column depth $y = E_b
(1+z)/4\pi R^2 Q_\mathrm{nuc}$, where $z$ is the redshift, $R$ the
neutron star radius and $Q_\mathrm{nuc} = 1.6 + 4 X$ MeV/nucleon the
nuclear energy release, given an H fraction X at ignition
\citep{Galloway06}. Assuming $z=0.31$ (the redshift for a 1.4
$M_\odot$, 10 km radius neutron star) we derive an ignition depth $y =
3.9\times 10^8$ g/cm$^2$ for solar abundances (X=0.7) and $y =
1.1\times 10^9 $ g/cm$^2$ for pure He (X=0). Ignition of pure He at
this column depth would require a heat flux from the deep crust of
$\gtrsim$ 2 MeV/nucleon at 1\% Eddington accretion rates (Figure 22,
\citealt{Cumming06}). This is higher than the crustal energy release
of $\sim$ 1.5 MeV/nucleon calculated by
\citet{haensel1990,haensel2003}, but close to the value of $\sim$ 1.9
MeV/nucleon recently found by \citet{gupta2007,haensel2008}. Recent
results by \citet{horowitz2008} demonstrate that even larger amounts
of heat may be released in the crust of an accreting neutron
star. However, the high heat requirements certainly put some strong
constraints on the pure He accretion scenario. Using the maximum local accretion
rate calculated above (1.9$\times 10^{3}$ g/cm$^2$/s), these ignition conditions
predict minimum recurrence times $\Delta t_\mathrm{rec} = y(1+z)/\dot{m}$ of
$\sim$10 days (X = 0) and $\sim$3 days (X = 0.7). The data provide
only weak constraints on the time without bursts that elapsed before
the long burst (we estimate that the source was in the field of view
of BAT for about 20\% of the two months analysed herein), but these
numbers are at least compatible with the time since the start of the
outburst.

\begin{table}
\center
\footnotesize
\caption{Properties of both type I X-ray bursts detected from XTE~J1701--407}
\begin{minipage}{\textwidth}
\begin{tabular}{ l c c}
\hline\hline
 & Long & Short \\
\hline
BAT rise time (s) & 50 & 5 \\
Duration\footnote{From start of rise to 1\% of peak flux.} (min) & 21 & ? \\
Peak flux\footnote{Bolometric} ($10^{-8}$ erg/s/cm$^2$) & 8.4 & $\sim$4.9  \\
Total fluence\footnote{Including fluence in BAT \citep{Markwardt08b} and \\ during XRT data gap.} ($10^{-6}$ erg/cm$^2$) & 3.5 & ? \\
Persistent flux\footnote{Unabsorbed 1-50~keV} ($10^{-10}$ erg/s/cm$^2$) & 8.6$\pm$0.9 & 9.8$\pm$1.0 \\
\hline\hline
\end{tabular}
\end{minipage}
\label{table:bursts}
\end{table}

The presence of a short, weaker burst at similar accretion rates is
perhaps the strongest piece of evidence pointing to mixed H/He fuel.
The short burst has a relatively slow rise time, $\approx 5$ s,
similar to that expected for mixed H/He bursts. The fact that it has
a lower peak flux than the long burst also implies some H content: if
the system were a pure He accretor we would expect both bursts to
reach similar peak fluxes since He bursts, whether long or short, are
expected to exhibit photospheric radius expansion \citep{Cumming04b}.
In the H-triggered burst scenario it is possible to have bursts with
quite different properties at very similar accretion rates, as H
burning transitions to stability \citep{Cooper07}.

XTE~J1701--407 increases the number of known NS-LMXBs in the
interesting low accretion rate bursting regime. The balance of
evidence suggests that it may be a good probe of thermonuclear burning
near the boundary of stable H ignition. However, further observations,
in particular better constraints on recurrence times and burst
energetics, are required to confirm this picture. Identification of
the companion star and determination of the orbital parameters may
also be able to confirm whether or not the system is ultracompact.

\section*{Acknowledgements}

ML would like to thank C. Markwardt and A. Cumming for stimulating
discussions and N. Rea, E. Mart{\' i}nez and Ll. Guasch for their
encouraging comments. RLCS acknowledges support from STFC.

\bibliographystyle{mn2e}

\end{document}